\documentclass[9pt,a4paper,twocolumn]{extarticle}
\usepackage{amsmath}
\usepackage[utf8]{inputenc}
\usepackage[T1]{fontenc}
\usepackage{XCharter}
\usepackage[xcharter,vvarbb]{newtxmath}
\usepackage{textcomp}
\usepackage{microtype}
\usepackage{graphicx}
\usepackage{booktabs}
\usepackage{tabularx}
\usepackage{array}
\usepackage{rotating}
\usepackage[table]{xcolor}
\usepackage{fancyhdr}
\usepackage{titlesec}
\usepackage[font=footnotesize,labelfont=bf,labelsep=period,justification=justified,singlelinecheck=false]{caption}
\usepackage[numbers,sort&compress]{natbib}
\definecolor{elslink}{RGB}{0,128,192}
\usepackage{multirow}
\usepackage[a4paper,top=22mm,bottom=20mm,left=16mm,right=16mm,columnsep=7mm,headsep=6mm]{geometry}
\usepackage[colorlinks,linkcolor=elslink,citecolor=elslink,urlcolor=elslink]{hyperref}

\titleformat{\section}{\normalfont\normalsize\bfseries}{\thesection.}{0.5em}{}
\titleformat{\subsection}{\normalfont\normalsize\itshape}{\thesubsection.}{0.5em}{}
\titleformat{\subsubsection}{\normalfont\normalsize\itshape}{\thesubsubsection.}{0.5em}{}
\titleformat{\paragraph}[runin]{\normalfont\normalsize\itshape}{}{0pt}{}[.]
\titlespacing*{\section}{0pt}{2.2ex plus .5ex minus .2ex}{1.2ex plus .2ex}
\titlespacing*{\subsection}{0pt}{1.8ex plus .5ex minus .2ex}{1ex plus .2ex}
\titlespacing*{\paragraph}{0pt}{0.8ex plus .3ex}{0.6em}
\fancypagestyle{plain}{\fancyhf{}\fancyfoot[C]{\footnotesize\thepage}}
\newcolumntype{L}{>{\raggedright\arraybackslash}X}
\begin{document}

\twocolumn[{%
\begin{@twocolumnfalse}
\vspace*{2mm}
{\color{black!70}\rule{\textwidth}{0.4pt}}\par\vspace{9mm}
{\fontsize{16}{20}\selectfont\raggedright Patient-wise evaluation of sleep-disorder classifiers on the CAP Sleep Database: a benchmark, record-wise inflation, and recording-system and age confounds\par}
\vspace{4mm}
{\large Azwad Jahin Ifat\textsuperscript{*,1}, Md.\ Sajjad Hossain Mridul\textsuperscript{1}, Mohammed Imamul Hassan Bhuiyan\par}
\vspace{2mm}
{\scriptsize\itshape Department of Electrical and Electronic Engineering, Bangladesh University of Engineering and Technology, Dhaka 1205, Bangladesh\par\textsuperscript{*}Corresponding author.\par\textsuperscript{1}These authors contributed equally to this work.\par\textit{E-mail addresses:} \textup{ifat.azwad@gmail.com}, \textup{Shmridul325@gmail.com}, \textup{imamul@eee.buet.ac.bd}\par}
\vspace{5mm}
\noindent\begin{minipage}[t]{0.27\textwidth}
{\scriptsize \textls[400]{\MakeUppercase{Article info}}}\par\vspace{-1.2mm}\rule{\linewidth}{0.4pt}\par\vspace{1mm}
{\scriptsize\textit{Keywords:}\\ sleep disorders\\ polysomnography\\ CAP Sleep Database\\ patient-wise cross-validation\\ data leakage\\ confounding\\ signal harmonisation\par}
\end{minipage}\hfill
\begin{minipage}[t]{0.69\textwidth}
{\scriptsize \textls[400]{\MakeUppercase{Abstract}}}\par\vspace{-1.2mm}\rule{\linewidth}{0.4pt}\par\vspace{1mm}
{\footnotesize Sleep disorders are diagnosed per patient. Yet automated classifiers built on the CAP Sleep Database report near-perfect accuracies for short segments under record-wise splits, in which segments of the same night appear in both training and test data. This study asks what performance remains when every test patient is unseen. Healthy controls and patients with insomnia, nocturnal frontal lobe epilepsy, periodic leg movements or REM sleep behaviour disorder were classified from harmonised central EEG, ECG and chin EMG. Each patient received one decision, under repeated patient-wise cross-validation and against empirical chance levels. A logistic regression of stage-wise feature aggregates set the patient-level benchmark at a balanced accuracy of 0.572%
. This was above chance but close to the 0.504 %
reached by age alone, and neither nested model selection nor a multimodal time--frequency convolutional network improved on it. Four published methods were then re-implemented and moved towards this protocol one safeguard at a time. Replacing their record-wise splits by patient-wise cross-validation lowered their segment-level balanced accuracy from 0.680%
--0.991 %
to 0.177%
--0.415 %
in every seed of every method. One decision per patient gave 0.331%
--0.447%
, below the benchmark. Record-wise splits inflated this study's own models in the same way. Recording system and age were entangled with diagnosis. Harmonisation lowered the area under the curve with which features identified the recording system among the healthy controls from 1.000 %
to 0.852%
, but did not remove the signal. Near-perfect results on this database are therefore consistent with recognition of individual recordings under record-wise splits.\par}
\vspace{1.5mm}\rule{\linewidth}{0.4pt}
\end{minipage}
\vspace{8mm}
\end{@twocolumnfalse}
}]

\section{Introduction}
\label{sec:intro}

Sleep is a regulated brain state that is vital to cognitive performance, health and well-being \cite{krueger2008}. Its timing and depth arise from the interaction of a homeostatic process, in which sleep pressure builds up during wakefulness, with a circadian process driven by the central pacemaker \cite{borbely2016}. Sleep consists of non-rapid eye movement (NREM) sleep, which is dominated by thalamocortical oscillations of the electroencephalogram (EEG), and rapid eye movement (REM) sleep, the state associated with dreaming \cite{hobson2002}. Both serve memory: slow-wave sleep, the deepest NREM stage, supports the consolidation of newly encoded memories, and subsequent REM sleep may stabilise them \cite{rasch2013}.

In polysomnography (PSG), EEG, eye movements, chin muscle tone (electromyography, EMG), leg movements and heart activity (electrocardiogram, ECG) are recorded overnight and scored in 30-s epochs. The NREM stages are distinguished by EEG rhythms, from theta waves through sleep spindles and K-complexes to the slow delta waves of deep sleep. REM sleep shows a wake-like EEG and atonia of the skeletal muscles \cite{carley2016,dauvilliers2018}.

Sleep disorders disturb this organisation in different ways. The International Classification of Sleep Disorders groups them into categories that include insomnia, sleep-related movement disorders and parasomnias \cite{sateia2014}. This study considers four disorder groups of the CAP Sleep Database, which differ in what PSG reveals about them. Insomnia makes sleep hard to initiate or maintain and is diagnosed mainly from the clinical history \cite{morin2015,littner2003}. Periodic leg movements (PLM) in sleep are a frequent PSG finding in several sleep disorders \cite{hornyak2006}. In REM sleep behaviour disorder (RBD), the atonia of REM sleep is lost \cite{dauvilliers2018}. Nocturnal frontal lobe epilepsy (NFLE), now termed sleep-related hypermotor epilepsy, causes seizures mainly during sleep that are hard to separate from parasomnias \cite{tinuper2016,tinuper2017}. The PSG signals carry complementary information about these groups. Central EEG holds the oscillations from which sleep stages are scored \cite{silber2007}. ECG reflects autonomic regulation through heart-rate variability, which has been studied in PLM and insomnia \cite{stein2012}. Chin EMG shows the REM muscle tone that is lost in RBD \cite{dauvilliers2018}.

Automated classifiers built on the openly available CAP Sleep Database \cite{terzano2001,capslpdb} report accuracies close to perfect. They range from handcrafted EEG features with tree ensembles \cite{sharma2021} and convolutional and recurrent networks on EEG or ECG segments \cite{wadichar2023,urtnasan2021} to multimodal fusion networks on EEG, ECG and EMG \cite{cheng2023,wang2024,subramaniam2025}. Most of them share three limitations. First, they score short segments or epochs, each labelled with the patient's diagnosis, instead of making one decision per patient. Second, most split the segments record-wise, so segments of every test patient's night are also seen in training. The few that hold out whole subjects still score epochs or pool several databases \cite{subramaniam2025,yaman2026}. Third, the results are not reported against an empirical chance level or together with the confounds of the database. Its recordings come from more than one acquisition system, and its diagnostic groups differ in age \cite{melnik2017,ohayon2004}. How much of the published performance would remain for unseen patients has therefore not been measured, nor how much of it depends on the evaluation design.

Patient-aware evaluation is needed because these disorders are diagnosed per patient. A model intended to support diagnosis will meet people whose recordings played no part in building it, and it should be judged on them. Record-wise splits do not answer this question. In a record-wise split (also called a segment-level split), the segments or epochs of each recording are assigned to folds independently. Because each patient contributes one recording, segments of the same night appear in both training and test data. The diagnosis is constant within a recording, so a model can succeed by recognising the recording, through the person's signals or the equipment, rather than the disorder. This form of leakage is known to inflate performance in clinical machine learning and in EEG deep learning (Section~\ref{sec:related}). Splitting by patient removes it. Making one decision per patient also matches the unit of diagnosis, so that patients with longer recordings do not weigh more.

This paper therefore addresses three aims. The first is to establish the performance achievable per patient when every test patient is unseen, and whether model complexity helps. The second is to measure how much the evaluation design inflates the reported performance of published methods. The third is to determine how far recording system and age explain the results. The main contributions are:
\begin{itemize}
\item \textbf{A patient-level benchmark:} A multi-class result on the CAP Sleep Database with one diagnosis per patient, under repeated patient-wise cross-validation and against an empirical chance level of the whole pipeline. To our knowledge, no reviewed CAP study reports this combination (Section~\ref{sec:related}). Neither a multimodal time--frequency convolutional neural network (CNN) nor nested model selection improves on handcrafted feature aggregates. The best result is above chance but close to age alone.
\item \textbf{A safeguard protocol for published methods:} A stepwise safeguard protocol that holds four published CAP classifiers fixed and adds one safeguard at a time: common cohort, task and channels; patient-wise split; harmonisation; and one decision per patient. With leave-one-safeguard-out checks, it locates where their reported performance is lost. For all four, most of it was lost at the split.
\item \textbf{A confound analysis:} Quantification of the recording-system and age confounds and of what harmonisation can remove. Every headline result is reported beside system-aware views, an age-only reference and contrasts between groups of similar age.
\end{itemize}

\section{Related works}
\label{sec:related}

Many automated sleep-disorder classifiers have been developed on the openly available CAP Sleep Database \cite{terzano2001,capslpdb}. Table~\ref{tab:lit} lists the reviewed studies whose full texts could be read, with their task, signals, model and unit scored. This section groups them by evaluation design. It moves from designs in which every test patient is also seen in training to designs that make one decision per unseen participant.

Most multi-class classifiers give the patient's diagnosis to short segments or epochs and split these record-wise. Segments of every test patient's night are then also in the training data. Wadichar et al.\ \cite{wadichar2023} classified 2-s segments of central EEG in this way with a convolutional recurrent network. Wang et al.\ \cite{wang2024} fused 30-s epochs of EEG, electro-oculography (EOG), ECG and EMG in an attention-based CNN. Cheng et al.\ \cite{cheng2023} reported the highest accuracy among these studies, 0.989%
. Their 90/10 split was made within each participant and sleep stage, so windows of every participant appear in both training and test data. Urtnasan et al.\ \cite{urtnasan2021} describe a split into 28 training and 7 test subjects for their single-lead ECG network. However, the per-class test fractions they report are consistent with a split of segments.

In other studies, the split is not stated. Sharma et al.\ \cite{sharma2021} evaluated their classifier of 30-s EEG epochs with 10-fold cross-validation, and Cui et al.\ \cite{cui2022} their insomnia and narcolepsy classifier with 7-fold cross-validation. Neither states whether the folds separated patients.

A third group holds out whole subjects but still scores short units, so the result is not a diagnosis per patient. Subramaniam et al.\ \cite{subramaniam2025} assigned each subject's recording to one of the training, validation and test sets. They also reported leave-one-subject-out validation, with one label per 30-s epoch. Sharma et al.\ \cite{sharma2021} added a 20\,\% subject hold-out. Philip Mulamoottil and Vigneswaran \cite{mulamoottil2024} held out one subject per cross-validation fold, on samples enlarged by two layers of synthetic generation.

Only a few studies make one decision per participant. Rechichi et al.\ \cite{rechichi2022} separated RBD from healthy controls with a chin-EMG atonia index and held out test subjects. In their cohort, however, the RBD patients were much older than the controls. Cooray et al.\ \cite{cooray2019} and Duarte et al.\ \cite{duarte2025} drew all or part of their healthy controls from other databases, so the source database differs between the classes. Duarte et al.\ also chose their threshold on all subjects. Two studies pooled CAP with other public databases and classified whole nights. Yaman et al.\ \cite{yaman2026} classified healthy sleep and five disorder classes from 15-s features with an attention-based fusion network. They kept the augmented copies of each subject on one side of every split, and reported lower accuracy under subject-independent than under subject-dependent cross-validation. They also noted that most earlier multi-class studies give every epoch the patient's diagnosis, and that many do not keep a patient's epochs out of the test set. Ali et al.\ \cite{ali2026} separated healthy from disordered sleep using the sleep-stage transitions of one hypnogram per subject. They found disorders more detectable in older than in younger subjects. Two further studies report subject-independent validation: a five-class CAP classifier validated in a second database \cite{alorf2026}, and an insomnia classifier of CAP-subphase measures evaluated leave-one-subject-out \cite{nagaiah2024}.

Why record-wise designs matter is well established. Record-wise cross-validation overestimates the accuracy of clinical prediction models \cite{saeb2017}. In EEG deep learning, segment-based hold-outs let subject-specific patterns leak \cite{brookshire2024}. Leakage is among the errors documented across machine-learning-based science \cite{kapoor2023}. More generally, models exploit any shortcut that separates the classes in the training data \cite{geirhos2020}. In medical imaging, networks have learned the hospital or the scanner instead of the disease \cite{zech2018,degrave2021}. A systematic review and a methodological review have found that such flaws leave few published models fit for clinical use \cite{roberts2021,varoquaux2022}.

Even when patients are separated, the CAP database carries confounds that a classifier can exploit. Its recordings come from more than one acquisition system, and EEG and ECG features depend on the recording system, the sampling rate and the filters \cite{melnik2017,widmann2015,ellis2015}. Its diagnostic groups also differ in age, and sleep architecture, sleep EEG power and heart-rate variability all change with age \cite{ohayon2004,carrier2001,umetani1998}.

To our knowledge, therefore, no reviewed study reports a multi-class CAP result with one diagnosis per patient under repeated patient-wise cross-validation, together with the recording-system and age confounds. None re-evaluated published CAP classifiers under patient-wise evaluation. Four of the methods in Table~\ref{tab:lit} were re-implemented in this study; the others are listed for their evaluation designs only.

\begin{table*}[t]
\centering\scriptsize\setlength{\tabcolsep}{3pt}\renewcommand{\arraystretch}{1.05}
\caption{Evaluation designs of the reviewed CAP Sleep Database classifiers, grouped as in Section~\ref{sec:related}. Record-wise split: segments of one recording can fall in training and test data; within-participant split: a record-wise split made within each participant.}
\label{tab:lit}
\begin{tabularx}{\textwidth}{@{}p{2.3cm}LLLLL@{}}
\toprule
\textbf{Study} & \textbf{Evaluation design} & \textbf{Unit scored} & \textbf{Classes} & \textbf{Signals} & \textbf{Model} \\
\midrule
\multicolumn{6}{@{}l}{\textit{Record-wise splits}} \\
Wadichar et al.\ \cite{wadichar2023} & 80/20 split and 10-fold CV of the extracted segments (record-wise) & 2-s segment & healthy vs disorder; five disorders & central EEG & convolutional recurrent network \\
Wang et al.\ \cite{wang2024} & 10-fold CV over epochs (record-wise) & 30-s epoch & healthy and five disorders & EEG, EOG, ECG, EMG & attention-fusion CNN \\
Cheng et al.\ \cite{cheng2023} & 90/10 split within each participant and stage, three repeats & 10-s window & all database groups & EEG, ECG, EMG spectrograms & VGG16 per signal, fusion network \\
Urtnasan et al.\ \cite{urtnasan2021} & one split described as 28 training and 7 test subjects (reported fractions consistent with a record-wise split) & 10-s segment (30\,s in the abstract) & healthy controls, insomnia, NFLE, PLM, RBD & single-lead ECG & 1-D CNN with recurrent layers \\
\multicolumn{6}{@{}l}{\textit{Split not stated}} \\
Sharma et al.\ \cite{sharma2021} & 10-fold CV (patient status of folds not stated); 20\,\% subject hold-out & 30-s epoch & healthy and six disorder groups & EEG (C4-A1, F4-C4) & wavelet Hjorth features, bagged trees \\
Cui et al.\ \cite{cui2022} & 7-fold CV (patient status of folds not stated) & 120-s window & healthy, insomnia, narcolepsy & ECG, respiration & random forest (CAP), support vector machine (diagnosis) \\
\multicolumn{6}{@{}l}{\textit{Subjects held out, short units scored}} \\
Subramaniam et al.\ \cite{subramaniam2025} & training, validation and test sets separated by subject; leave-one-subject-out & 30-s epoch & healthy and five disorders & EEG, EOG, EMG & 1-D CNN with multimodal transformer encoder \\
Philip Mulamoottil and Vigneswaran \cite{mulamoottil2024} & 5-fold CV; CV holding out one subject per fold & augmented sample & insomnia vs healthy & single-channel EEG & ensemble and tree-based classifiers \\
\multicolumn{6}{@{}l}{\textit{One decision per participant}} \\
Rechichi et al.\ \cite{rechichi2022} & CV on a training set; held-out test subjects & participant & RBD vs healthy controls & chin EMG & $k$-nearest neighbours, support vector machine \\
Cooray et al.\ \cite{cooray2019} & 10-fold CV; healthy controls from another database & participant & RBD vs healthy controls & EEG, EOG, EMG & random forest \\
Duarte et al.\ \cite{duarte2025} & threshold chosen on all subjects; healthy controls partly from another database & participant & disorders vs healthy controls & one EEG channel & threshold on a fractal scaling exponent \\
Yaman et al.\ \cite{yaman2026} & subject-independent hold-out and 4-fold CV; three pooled databases & recording & healthy and five disorder classes & EEG, EOG, chin and leg EMG, ECG & attention-based multimodal fusion network \\
Ali et al.\ \cite{ali2026} & 5-fold CV of subjects, stratified by database and disorder; four pooled databases & participant (hypnogram) & healthy vs disordered & hypnogram (stage transitions) & tree ensembles \\
\midrule
This study & repeated stratified patient-wise 5-fold CV & patient & healthy and four disorders & central EEG, ECG, chin EMG & handcrafted features with four classifiers; multimodal CNN \\
\bottomrule
\end{tabularx}
\end{table*}

\section{Materials and methods}
\label{sec:methods}

\begin{figure*}[t]
\centering
\includegraphics[width=\textwidth]{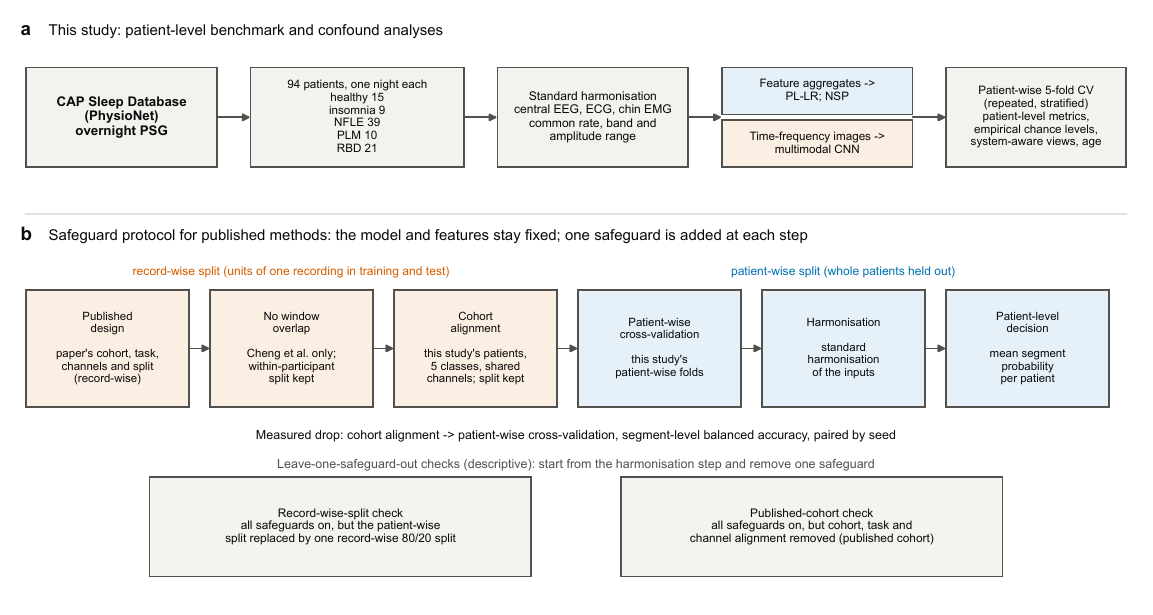}
\caption{Study overview. (a) Patient-level benchmark and confound analyses. (b) Safeguard protocol for the published methods: each step adds one safeguard with model and features fixed; the leave-one-safeguard-out checks each remove one safeguard from the harmonisation step.}
\label{fig:pipeline}
\end{figure*}

Fig.~\ref{fig:pipeline} gives an overview of the benchmark and of the safeguard protocol.

\subsection{Dataset and cohort}
\label{sec:m-data}

We used the CAP Sleep Database v1.0.0 on PhysioNet \cite{capslpdb,pollard2026}, collected at the Sleep Disorders Center of the Ospedale Maggiore of Parma, Italy \cite{terzano2001}. It provides overnight PSG of healthy controls and several disorder groups. It also provides expert annotations of the cyclic alternating pattern (CAP), an EEG marker of unstable non-REM sleep \cite{parrino2012}. According to the database documentation, expert neurologists scored sleep stages under the Rechtschaffen and Kales rules \cite{rechtschaffen1968} as wake, NREM stages S1 to S4, REM sleep and movement time. The healthy controls had no neurological disorder and took no drugs affecting the central nervous system. Stages S3 and S4 together correspond approximately to stage N3 of the current AASM rules \cite{silber2007}. We grouped the stages into wake, light sleep (S1 and S2), deep sleep (S3 and S4) and REM sleep. Diagnoses were those provided with the database; patients were not re-diagnosed. We analysed five groups: healthy controls, insomnia, NFLE (the database label), PLM and RBD. Narcolepsy, sleep-disordered breathing and bruxism were not analysed, because their groups were too small for patient-wise cross-validation.

All 108 %
recordings were audited before analysis. Apart from the three groups not analysed, three recordings were excluded. One was a healthy-control recording with EEG only, one an NFLE recording whose hypnogram starts well before the signal, and one an RBD recording that duplicates the signal file of another RBD patient. The cohort comprised 94 %
patients (Table~\ref{tab:cohort}). Hypnograms were aligned to the signal start, and 30-s epochs not entirely within the signal were dropped. Age and sex came from the database metadata; the youngest patients were adolescents in the NFLE group (Table~\ref{tab:cohort}). Age was never a model input except in the labelled age analyses.

The cohort spans more than one recording system. Each recording was assigned from its signal-file header to the main recording system or to other recording systems, which differ in sampling rate, hardware filters and amplitude range. The recording system was never a model input. The other systems held mostly healthy controls (Table~\ref{tab:cohort}), so a main-system-only cohort would have kept only 6 %
healthy controls. All patients were therefore kept, and every headline result is reported next to pre-specified system-aware views.

\begin{table*}[t]
\centering\footnotesize\setlength{\tabcolsep}{4pt}
\caption{Cohort by class (database metadata and signal-file headers).}
\label{tab:cohort}
\begin{tabular}{@{}lcccccc@{}}
\toprule
\textbf{Class} & \textbf{Patients} & \textbf{Median age} & \textbf{Age range} & \textbf{Women} & \textbf{Main} & \textbf{Other} \\
 & & \textbf{(years)} & \textbf{(years)} & & \textbf{system} & \textbf{systems} \\
\midrule
Healthy controls & 15 %
& 31.0 %
& 23%
--42 %
& 8 %
& 6 %
& 9 %
\\
Insomnia & 9 %
& 58.0 %
& 47%
--82 %
& 5 %
& 9 %
& 0 %
\\
NFLE & 39 %
& 29.0 %
& 14%
--67 %
& 18 %
& 37 %
& 2 %
\\
PLM & 10 %
& 58.0 %
& 40%
--62 %
& 3 %
& 10 %
& 0 %
\\
RBD & 21 %
& 72.0 %
& 58%
--82 %
& 3 %
& 21 %
& 0 %
\\
\midrule
All & 94 %
& & & & 83 %
& 11 %
\\
\bottomrule
\end{tabular}
\end{table*}

\subsection{Data preprocessing}
\label{sec:m-harm}

Three signals were used: central EEG (C4-A1, or C3-A2 when C4-A1 was unavailable), ECG and chin EMG. Leg EMG was not used because its availability differs by class. Preprocessing aimed to stop properties of the recording system from acting as class cues. It mapped every recording to one representation that the lowest-rate recording can support (the standard harmonisation). EEG was converted to microvolts and clipped at $\pm$50\,\textmu V, the narrowest amplitude range in the cohort, so that saturation is the same in every recording. All signals were then resampled to 128\,Hz with anti-aliased polyphase resampling. They were band-pass filtered with a fourth-order Butterworth filter applied forwards and backwards (zero phase; 0.5--25\,Hz for EEG and ECG, 10--45\,Hz for chin EMG). Each filtered signal $x$ was robust-z-scored per recording,
\begin{equation}
z = \frac{x - \mathrm{median}(x)}{\mathrm{IQR}(x)/1.349},
\label{eq:robustz}
\end{equation}
a z-score in which the median and the interquartile range (IQR) replace the mean and the standard deviation. Both are computed over the scored period of the recording (the epochs covered by the hypnogram). Dividing by 1.349, the IQR of a standard normal distribution, gives $z$ unit standard deviation for Gaussian data. The score is insensitive to artefacts and to the amplitude scale of the recording system. Epochs were removed only when a signal was flat or of extreme amplitude. Two stricter harmonisations, a filter-matched and a spectrally equalised one (Section~\ref{sec:r-confounds}), were evaluated before any deep-learning data were produced. Neither lowered the recording-system signal, so the standard harmonisation was used throughout.

\subsection{Models}
\label{sec:m-models}

\begin{table*}[t]
\centering\scriptsize\setlength{\tabcolsep}{3pt}
\caption{Candidate pipelines of the nested selection procedure (NSP). PL-LR: patient-level logistic regression; aggregates: patient-level feature aggregates; biomarkers: whole-night biomarkers.}
\label{tab:nested}
\begin{tabularx}{\textwidth}{@{}p{0.6cm}>{\raggedright\arraybackslash}p{3.4cm}>{\raggedright\arraybackslash}p{4.2cm}L@{}}
\toprule
\textbf{ID} & \textbf{Model} & \textbf{Inputs} & \textbf{Relation to the PL-LR} \\
\midrule
C1 & L2 logistic regression & aggregates & the PL-LR, re-tuned inside the NSP \\
C2 & L2 logistic regression & aggregates + biomarkers & same model, more features \\
C3 & L2 logistic regression & aggregates + biomarkers, equipment-sensitive features removed & same model, filtered features \\
C4 & elastic-net logistic regression & aggregates + biomarkers & still logistic regression, but with a combined L1 + L2 penalty, so that it can drop features \\
C5 & two logistic regressions in sequence & aggregates + biomarkers & different structure: healthy versus disorder first, then the disorder \\
C6 & gradient-boosted trees (LightGBM) & aggregates + biomarkers & not logistic regression: a non-linear tree ensemble \\
C7 & average of the C2 and C6 probabilities & aggregates + biomarkers & half logistic regression, half trees \\
\bottomrule
\end{tabularx}
\end{table*}

\begin{figure*}[t]
\centering
\includegraphics[width=\textwidth]{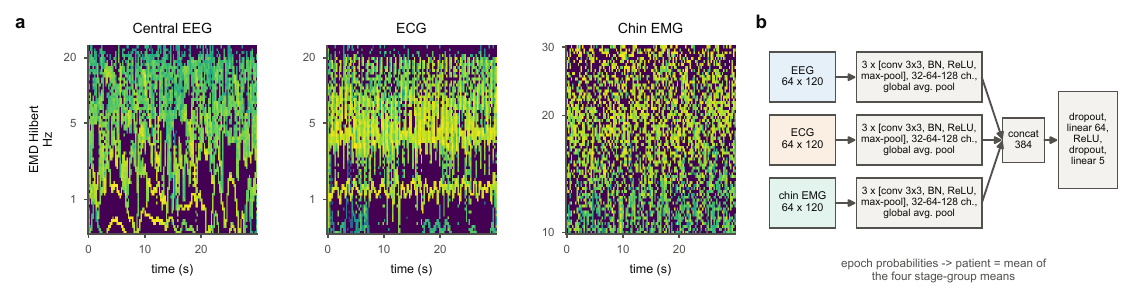}
\caption{Time--frequency inputs and the multimodal CNN. (a) One 30-s stage-2 epoch of a healthy control as EMD Hilbert spectra, the selected representation, for each signal: a $64 \times 120$ log-power image with 64 log-spaced frequencies (0.5--25\,Hz for EEG and ECG, 10--30\,Hz for chin EMG) and 120 time steps of 0.25\,s; colour scale per panel, 2nd--99.5th percentile of log$_{10}$ power. (b) The multimodal CNN, on EMD Hilbert spectra: one convolutional branch per signal, concatenation and a two-layer head; the patient decision is the mean of the four stage-group mean epoch probabilities.}
\label{fig:tf}
\end{figure*}

\paragraph{Feature aggregates and the patient-level logistic regression (PL-LR)} The benchmark model summarises each patient's night by stage-wise feature aggregates. For each 30-s epoch we computed EEG band powers (delta to low beta, absolute and relative), spectral entropy, spectral edge frequency and Hjorth parameters. From ECG R-peaks we computed heart rate and heart-rate variability (SDNN, RMSSD, LF/HF), and from chin EMG tone and phasic-activity indices. Per patient, each epoch feature $f_j$ was summarised within each of the four stage groups $g$ of the expert hypnogram (wake, light sleep S1--S2, deep sleep S3--S4, REM) by
\begin{equation}
\mathbf{a}_{j,g} = \bigl(Q_{50},\; Q_{75}-Q_{25},\; Q_{10},\; Q_{90}\bigr)\bigl(\{f_j(e) : e \in E_g\}\bigr),
\label{eq:agg}
\end{equation}
where $E_g$ is the set of the patient's epochs in stage group $g$ and $Q_q$ denotes the $q$th percentile (median, IQR, 10th and 90th percentiles). The concatenation of all $\mathbf{a}_{j,g}$ is the patient's vector of feature aggregates $\mathbf{x}$. Features describing the hypnogram or the CAP annotations were excluded from all primary models. Missing values, such as a stage group absent in a patient, were imputed. Four classifiers were fitted on the feature aggregates: L2-regularised multinomial logistic regression, random forest, gradient-boosted trees and a radial-basis support vector machine. Imputation and scaling were fitted on training patients, and hyperparameters were tuned in an inner stratified 3-fold search. The logistic regression had the highest mean balanced accuracy and was carried forward as the \emph{patient-level logistic regression} (PL-LR). For a patient with standardised feature aggregates $\tilde{\mathbf{x}}$, it gives the probability of class $c \in \{1, \dots, K\}$, $K = 5$, as
\begin{equation}
p(c \mid \tilde{\mathbf{x}}) = \frac{\exp(\mathbf{w}_c^{\top}\tilde{\mathbf{x}} + b_c)}{\sum_{k=1}^{K} \exp(\mathbf{w}_k^{\top}\tilde{\mathbf{x}} + b_k)},
\label{eq:lr}
\end{equation}
with weights fitted on the $N$ training patients by minimising the class-weighted, L2-penalised cross-entropy
\begin{equation}
\mathcal{L}(\mathbf{W}, \mathbf{b}) = \frac{1}{2}\sum_{k=1}^{K}\lVert\mathbf{w}_k\rVert_2^2 - C\sum_{i=1}^{N} \omega_{y_i} \log p(y_i \mid \tilde{\mathbf{x}}_i), \qquad \omega_c = \frac{N}{K N_c},
\label{eq:lrloss}
\end{equation}
where $N_c$ is the number of training patients of class $c$, so that every class contributes equally. The inverse regularisation strength $C$ was chosen from $\{0.01, 0.1, 1\}$ in the inner search. The patient is assigned the class of highest probability. Because the PL-LR was chosen on the same cross-validation data, its estimate is optimistic; the nested selection procedure (NSP, below) re-estimates it without that bias.

\paragraph{Nested selection procedure (NSP)} The NSP tests whether a better patient-level pipeline exists, without the optimism of choosing it on the same data. Seven candidate pipelines (C1--C7, Table~\ref{tab:nested}) were fixed in advance. Each keeps the patient-level inputs and varies one aspect of the PL-LR: the features (C2, C3), the penalty (C4), the structure of the task (C5) or the model class (C6, C7). C1 is the PL-LR itself. C2 adds whole-night biomarkers, intended to raise the recall of insomnia and PLM, the weakest disorder classes. These are spindle and slow-wave measures detected with the YASA toolbox \cite{vallat2021}, a relative REM atonia index of chin EMG \cite{ferri2010}, relative low-beta power and heart-rate measures. C3, the system-filtered candidate, drops features whose healthy-control recording-system area under the ROC curve (AUC) exceeds 0.75, estimated inside each training set. C4 is an elastic-net logistic regression, whose combined L1 and L2 penalty can set the weights of uninformative features to zero. C5 is a two-stage model. A calibrated healthy-versus-disorder logistic regression gives the probability of healthy, $p_1$, and a four-class disorder logistic regression divides the remaining $1 - p_1$ among the disorders. C6 is a gradient-boosted tree ensemble (LightGBM), which can capture non-linear effects and interactions between features. C7 is a soft vote that averages the class probabilities of C2 and C6, each at its own inner-selected setting. The candidates were compared by nested cross-validation, with 4 folds $\times$ 3 repeats inside each outer training set. Every selection, tuning and calibration step was fitted on training patients only \cite{varma2006,cawley2010}. A dedicated, calibrated healthy-versus-disorder model was evaluated in the same way. Its pre-specified screening operating point required a sensitivity of at least 0.8 and a specificity of at least 0.5.

\paragraph{Deep-learning comparator} To test whether a more flexible model improves on the PL-LR, a multimodal CNN was trained on the same patients. One hundred quality-checked epochs per patient, 25 from each stage group, were selected by a deterministic time-stratified rule. Each epoch of each signal was converted into a $64 \times 120$ log-power time--frequency image. Four representations were compared: the short-time Fourier transform (STFT), the continuous wavelet transform (CWT), and the Hilbert spectra of empirical mode decomposition (EMD) \cite{huang1998} and multivariate EMD (MEMD) \cite{rehman2010}. A raw-signal 1-D network was a further comparator. Each signal has its own convolutional branch, and the branches are concatenated before a classification head (Fig.~\ref{fig:tf}). The network was trained on epochs with the class-weighted cross-entropy of Eq.~\eqref{eq:lrloss} and mixup. The patient's class probabilities were the mean of the four stage-group mean epoch probabilities. This rule was fixed before any CNN run, because wake availability differs by class. The representation, fusion and ablations were chosen on inner-validation balanced accuracy only. The selected configuration (EMD Hilbert spectra, concatenation, all three signals) is the \emph{multimodal CNN}.

\subsection{Evaluation protocol}
\label{sec:m-eval}

All patient-level results used stratified patient-wise 5-fold cross-validation repeated with seeds 0--4. The folds of a seed were shared by every analysis, so results are paired by seed. Metrics were computed on the pooled out-of-fold predictions of each seed. The primary metric was the 5-class balanced accuracy (mean per-class recall). Macro-F1, Cohen's $\kappa$ and per-class recall are also reported, and for healthy versus disorder the AUC of the disorder probability (one minus the healthy probability). Confidence intervals (CIs) are 95\,\% patient-level bootstrap intervals (2000 resamples) computed per seed. Where a value is a mean over seeds, the reported interval is the range of the per-seed intervals, from the lowest lower bound to the highest upper bound. The range was used because the seeds re-partition the same patients, so their intervals are not independent estimates that could be pooled. It is the conservative summary that contains the interval of every seed. It has no stated coverage and, because it resamples predictions of fixed runs, excludes model-fitting variance. The empirical chance level was estimated by refitting the whole pipeline with training labels permuted within training folds, 100 times per seed, for the PL-LR and for the NSP \cite{ojala2010}. A learning curve of the PL-LR inside the nested protocol used growing shares of each outer training set.

Pre-specified system-aware views were reported regardless of outcome. They comprise balanced accuracy on main-system patients only, healthy recall by system, and the AUC with which inputs or model outputs separate the two systems within the healthy controls (healthy-control system AUC). A fourth view is a 4-class disorder-only endpoint on main-system patients, with the healthy column dropped and the disorder probabilities renormalised. It removes the recording-system confound but not age. It was specified before the CNN results were seen; for the PL-LR, it was applied to stored predictions after the 5-class results were known. To measure inflation within this study, three models were also evaluated with record-wise splits of epochs on identical inputs: epoch-level gradient-boosted trees, the multimodal CNN and an epoch-level logistic regression. Under a record-wise split every patient has epochs in the training data, so a patient-level decision would always concern a partly seen patient. These comparisons were therefore made at the epoch level. Outside all test families, we report three pre-specified descriptive age analyses: an age-only multinomial model, the feature aggregates added to age, and insomnia versus PLM on the main system, two groups of similar age. One post hoc contrast compares healthy controls with NFLE on all systems.

\subsection{Replication and safeguard protocol}
\label{sec:m-safeguards}

\paragraph{Replications} Four published CAP classifiers were re-implemented from their papers \cite{sharma2021,wadichar2023,urtnasan2021,cheng2023}, chosen because their protocols were described in enough detail to be rebuilt. Each was first run at its \emph{published design}, using the paper's own cohort, task, channels, preprocessing and split. It was compared with the published result on the authors' metric (Table~\ref{tab:replication}). Gaps in a paper were filled from the authors' public code where available, otherwise by a standard choice recorded before any run. No setting was changed afterwards to approach a published number. Three deviations and reporting inconsistencies shape the comparisons. (i) The optimised filter bank of Sharma et al.\ could not be rebuilt, because its coefficients are not published. A biorthogonal wavelet decomposition replaced it, and this replication is called Sharma-style. (ii) Cheng et al.'s VGG16 networks were replaced by a VGG-style network of quarter width, for compute reasons. (iii) Urtnasan et al.\ describe a split into training and test subjects, but the per-class test fractions they report are consistent with a split of segments. We therefore infer that their published result came from a record-wise segment split, and used such a split at the published design and at cohort alignment. The main comparison, of the same model under a record-wise and a patient-wise split, does not depend on this inference; only the link to the published number does. ECG was converted to millivolts before this re-implementation was carried forward, because the recordings store ECG in different units.
\paragraph{Safeguard steps} The published methods were evaluated in five steps. Each step kept everything from the previous one and changed one thing, while the model, features and training recipe stayed as published (Fig.~\ref{fig:pipeline}b):
\begin{enumerate}
\item \textbf{Published design:} Each paper's own cohort, task, channels, preprocessing and split, to check that the re-implementation reproduces the published result.
\item \textbf{Cohort alignment:} This study's patients, five classes and channels, with the record-wise split of the published design kept. This step is not itself a protection against leakage; it makes the steps comparable, because all methods are then evaluated on the same data.
\item \textbf{Patient-wise cross-validation:} The record-wise split replaced by this study's patient-wise 5-fold folds, so that no test patient contributes segments to training. Each segment is still scored separately.
\item \textbf{Harmonisation:} The standard harmonisation of the inputs, to reduce differences between recording systems.
\item \textbf{Patient-level decision:} The segment probabilities averaged per patient, giving one diagnosis per patient.
\end{enumerate}
Recordings below a method's native sampling rate were upsampled to that rate at cohort alignment. The sampling-rate cue was therefore present by design until harmonisation, and equally in the two steps compared to measure the drop. For Cheng et al., an additional step between the published design and cohort alignment removed the overlap of the 10-s windows. A thinned-window control separates window overlap from training-set size. The first four steps score each segment separately (for Cheng et al., each window), so their values are segment-level even after the split becomes patient-wise. Only the fifth step scores each patient. Segment-level metrics pool all test segments of a fold, so patients with more segments weigh more. Each method's own unit is used: 2-s or 10-s segments, 30-s epochs, or for Cheng et al.\ 10-s windows, reported as window level. Two descriptive leave-one-safeguard-out checks started from the harmonisation step and each removed one safeguard. The record-wise-split check replaced the patient-wise split by one stratified 80/20 split of segments. The published-cohort check removed the cohort, task and channel alignment (for Cheng et al., post hoc and on chin EMG only). A third check, post hoc and for Wadichar et al.\ and the Sharma-style replication only, applied the split of the record-wise-split check to the native signals of cohort alignment. It was added after the leave-one-safeguard-out results were known, to separate the effect of harmonisation under a record-wise split from the change of split.

\paragraph{Drop under patient-wise splits} If a published result depends on segments of the test patients also appearing in training, its segment-level balanced accuracy should fall when the record-wise split of cohort alignment is replaced by patient-wise cross-validation. For Cheng et al., the measure is the window-level balanced accuracy of the fused decision network. This drop was measured in every replication, paired by seed (seeds 0--2). Before the runs, a drop was defined as substantial only if it was at least 0.05 in every seed and on average in every replication. A one-sided paired t-test per replication, Holm-corrected across the four, was secondary. With so small a margin, the size of the drop is more informative than whether the margin was met.

\begin{table*}[t]
\centering\scriptsize\setlength{\tabcolsep}{2.5pt}\renewcommand{\arraystretch}{1.1}
\caption{Published results, replications of the published designs and the same methods under the safeguards. Published: accuracy as reported (Urtnasan et al.: per-class F1 range). Replication: accuracy (Urtnasan et al.: median macro-F1 over fifteen subject draws). Safeguard columns: balanced accuracy, mean over the three seeds (Cheng et al.: window level); cohort alignment keeps the record-wise split. This study: patient-level balanced accuracy on the same folds, mean over the five seeds (PL-LR: patient-level logistic regression; NSP: nested selection procedure). CP2: Wadichar et al.'s five-disorder task.}
\label{tab:replication}
\begin{tabularx}{\textwidth}{@{}>{\raggedright\arraybackslash}p{1.9cm}LL*{4}{>{\centering\arraybackslash}p{1.05cm}}@{}}
\toprule
& & & \multicolumn{3}{c}{\textbf{Segment level}} & \textbf{Patient} \\
\cmidrule(lr){4-6}\cmidrule(l){7-7}
\textbf{Method} & \textbf{Published} & \textbf{Replicated published design} & \textbf{Cohort align.} & \textbf{Patient-wise CV} & \textbf{Harmo\-nisation} & \textbf{Decision} \\
\midrule
Wadichar et al.\ \cite{wadichar2023} & CP2: CV 0.831%
; hold-out 0.905 %
& CP2: CV 0.831%
; record-wise hold-out 0.765 %
& 0.811 %
& 0.177 %
& 0.175 %
& 0.331 %
\\
Sharma et al.\ \cite{sharma2021} (Sharma-style) & CV 0.913%
; subject hold-out 0.907 %
& CV 0.829%
; subject hold-out 0.476 %
& 0.680 %
& 0.315 %
& 0.324 %
& 0.332 %
\\
Cheng et al.\ \cite{cheng2023} & fusion 0.989 %
(VGG16) & within-participant split 0.9997 %
(quarter-width network; one repeat) & 0.983 %
& 0.415 %
& 0.371 %
& 0.414 %
\\
Urtnasan et al.\ \cite{urtnasan2021} & per-class F1 0.950%
--0.990 %
& record-wise segment split 0.995%
; subject split 0.256 %
& 0.991 %
& 0.390 %
& 0.395 %
& 0.447 %
\\
\midrule
This study, PL-LR & --- & --- & --- & --- & --- & 0.572 %
\\
This study, NSP & ---& --- & --- & --- & --- & 0.572 %
\\
\bottomrule
\end{tabularx}
\end{table*}

\subsection{Explainability and statistical analysis}
\label{sec:m-stats}

Each fold model of the PL-LR was explained with exact linear SHAP values on its own test patients \cite{lundberg2017}. We first confirmed that the refitted fold models reproduced every original out-of-fold prediction. We report per-class attributions summed by signal group, and the 15 leading features with their correlation with age and their healthy-control system AUC. We also report the stability of rankings across folds and seeds. SHAP values describe what the model uses, not mechanisms of disease \cite{ghassemi2021}.

Each analysis stage and the safeguard analysis had a pre-specified confirmatory family of tests, with Holm correction across the family at $\alpha = 0.05$. Pipelines were compared by paired t-tests on the seed-level balanced accuracies. Tests were two-sided, or one-sided where the direction was fixed before the runs, for example whether a more complex model was better. Because seeds share patients, these tests measure variation across cross-validation repetitions, not across patients. Repeated resampling also inflates their type I error \cite{dietterich1998}. The p-values are therefore descriptive, not inference about the patient population. Conclusions rest on the effect sizes (mean paired differences) and seed-level values, judged against criteria fixed before the runs. Descriptive and post hoc analyses are reported outside the families.

\section{Results}
\label{sec:results}

\subsection{Patient-level benchmark}
\label{sec:r-benchmark}

The results follow the argument of the paper. This section sets the patient-level benchmark. Sections~\ref{sec:r-published} and~\ref{sec:r-inflation} ask why published results on this database are much higher, first for four published methods and then for this study's own models. Section~\ref{sec:r-confounds} asks how much of the benchmark itself is confounded by recording system and age, and Section~\ref{sec:r-shap} closes with what the benchmark model uses.

The PL-LR reached a 5-class balanced accuracy of 0.572 %
under patient-wise cross-validation (Table~\ref{tab:main}). The empirical chance level, the mean of the 100-permutation null of the same pipeline, was 0.199%
. Every seed exceeded its permutation null (largest empirical $p = 0.010$%
, the smallest attainable). Age alone, however, reached 0.504 %
(Table~\ref{tab:main}). The benchmark is therefore above chance but close to what age achieves; Section~\ref{sec:r-confounds} examines this confound. The NSP, which removes the optimism of choosing the PL-LR on the same data, reached the same value (0.572%
). Every seed again exceeded the permutation null of the whole procedure at the smallest attainable p. The inner loops chose the PL-LR (C1) in 68\,\% %
of outer folds, and the NSP did not exceed the PL-LR (mean paired difference 0.000%
, Holm $p = 1.000$%
).

More complex models did not improve on the PL-LR. The multimodal CNN reached 0.388%
, a mean paired difference of -0.184 %
from the PL-LR (Holm $p = 1.000$%
, one-sided test of whether the CNN was better). With the handcrafted features added, it reached 0.574%
, the level of the PL-LR alone. No pre-specified variant changed its balanced accuracy detectably: not the time--frequency representation, multivariate EMD, attention fusion, signal subsets or network width. With five seeds, this is not evidence that they are equivalent.

\begin{figure}[tb]
\centering
\includegraphics[width=\columnwidth]{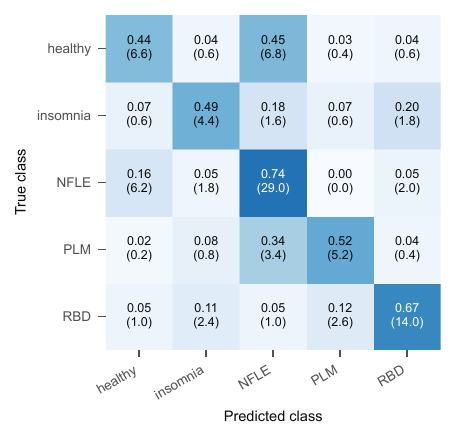}
\caption{Confusion matrix of the patient-level logistic regression (PL-LR; evaluated directly, not inside the nested selection procedure) under patient-wise 5-fold cross-validation: row proportions, mean patient counts over the five seeds in brackets; the diagonal is the per-class recall.}
\label{fig:confusion}
\end{figure}

The errors fell mostly on healthy controls and insomnia. Recall was highest for NFLE (0.744%
) and RBD (0.667%
) and lowest for healthy controls (0.440%
) and insomnia (0.489%
). Healthy controls were most often classified as NFLE (Fig.~\ref{fig:confusion}). Whole-night biomarkers did not raise insomnia and PLM recall (difference 0.000%
). Healthy-versus-disorder discrimination was also weak. The PL-LR ranked patients with AUC 0.800 %
but had a specificity of 0.267 %
at the default threshold. A dedicated calibrated model reached a mean sensitivity of 0.772 %
and specificity of 0.533%
. No seed met both targets of the pre-specified screening operating point.

The PL-LR therefore sets the benchmark. Published CAP classifiers report accuracies of up to 0.989 %
\cite{cheng2023}, mostly per segment and under record-wise splits. Section~\ref{sec:r-published} tests whether this difference lies in the methods or in their evaluation.

\begin{table*}[t]
\centering\footnotesize\setlength{\tabcolsep}{3pt}
\caption{Patient-level results under patient-wise 5-fold cross-validation (94 %
patients; mean over the five seeds). PL-LR: patient-level logistic regression; NSP: nested selection procedure. CI: range of the per-seed patient-bootstrap intervals. Empirical chance level: null distribution of the PL-LR pipeline with labels permuted 100 times per seed; values are null means over seeds, and the interval runs from the lowest 2.5th to the highest 97.5th percentile of the per-seed nulls ($\kappa$ was not stored for the permutations). Age rows: $\kappa$ and CIs computed afterwards from out-of-fold predictions (age + feature aggregates: from a re-run that reproduced the original values exactly).}
\label{tab:main}
\begin{tabularx}{\textwidth}{@{}Lcccccc@{}}
\toprule
& \multicolumn{4}{c}{\textbf{5 classes}} & \multicolumn{2}{c}{\textbf{Healthy vs disorder}} \\
\cmidrule(lr){2-5}\cmidrule(l){6-7}
\textbf{Model} & \textbf{Bal.\ acc.} & \textbf{CI} & \textbf{Macro-F1} & \textbf{$\kappa$} & \textbf{AUC} & \textbf{CI} \\
\midrule
Empirical chance level (PL-LR pipeline, 100 permutations per seed) & 0.199 %
& 0.107%
--0.300 %
& 0.197 %
& --- & 0.498 %
& 0.338%
--0.656 %
\\
Age only & 0.504 %
& 0.324%
--0.651 %
& 0.502 %
& 0.448 %
& 0.709 %
& 0.592%
--0.817 %
\\
PL-LR & 0.572 %
& 0.433%
--0.736 %
& 0.577 %
& 0.490 %
& 0.800 %
& 0.574%
--0.950 %
\\
NSP & 0.572 %
& 0.423%
--0.753 %
& 0.573 %
& 0.487 %
& 0.806 %
& 0.595%
--0.938 %
\\
Age + feature aggregates & 0.572 %
& 0.433%
--0.753 %
& 0.576 %
& 0.490 %
& 0.819 %
& 0.599%
--0.964 %
\\
Multimodal CNN & 0.388 %
& 0.277%
--0.507 %
& 0.369 %
& 0.265 %
& 0.724 %
& 0.497%
--0.914 %
\\
\bottomrule
\end{tabularx}
\end{table*}

\subsection{Published methods under the safeguards}
\label{sec:r-published}

Each re-implementation was first checked against its paper and then taken through the five safeguard steps (Section~\ref{sec:m-safeguards}). At their published designs (step 1; Table~\ref{tab:replication}), the re-implementations matched or approached the published values under record-wise splits. Wadichar et al.\ reached 0.831 %
against 0.831 %
with 10-fold cross-validation. Urtnasan et al.\ reached a median macro-F1 of 0.995 %
with a record-wise segment split, at or above the published per-class F1 values, which are not directly comparable. Cheng et al.\ reached 0.9997 %
against 0.989 %
with its within-participant split (one repeat). The Sharma-style re-implementation fell short with epoch-level cross-validation (0.829 %
against 0.913%
). This is plausibly due in part to the substituted filter bank, whose coefficients are not published. Where whole subjects were held out, values were far lower: 0.256 %
for Urtnasan et al.'s subject split, and 0.476 %
for the Sharma-style subject hold-out against a published 0.907%
. That published result is the only one with subjects held out. From these data we cannot tell whether our shortfall reflects leakage in the original evaluation or the weaker re-implementation. Wadichar et al.'s hold-out (0.765 %
against 0.905%
) is itself a record-wise split and says nothing about unseen people.

\begin{figure*}[t]
\centering
\includegraphics[width=\textwidth]{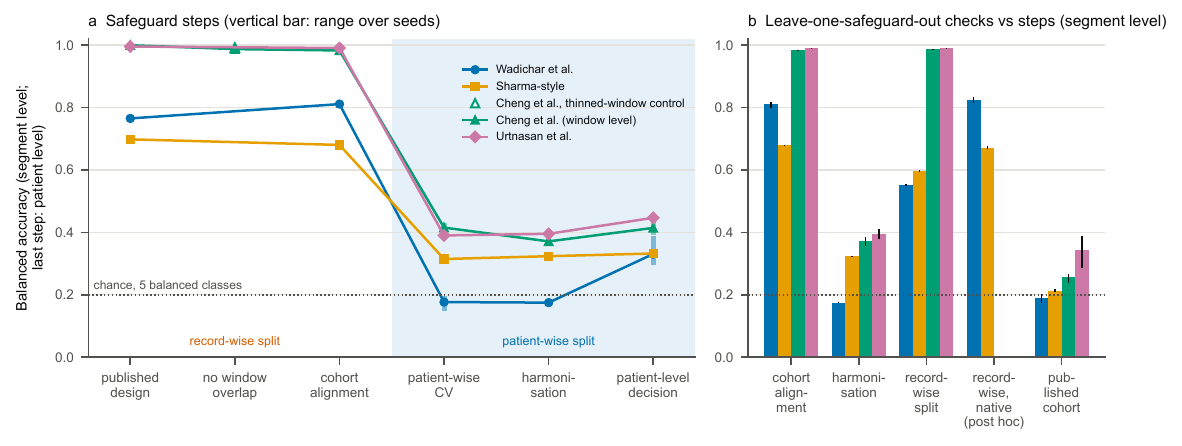}
\caption{Published methods under the safeguard protocol. (a) Balanced accuracy at each step, segment level (Cheng et al.: window level) up to harmonisation and patient level at the last step. Published design: each paper's own class set and split (Wadichar et al.: five-disorder task, record-wise hold-out; Sharma-style: 10-fold cross-validation; Urtnasan et al.: median of fifteen subject draws, record-wise segment split inferred from the reported fractions); accuracies in Table~\ref{tab:replication}. Open marker: Cheng et al.'s thinned-window control. (b) Leave-one-safeguard-out checks beside the step they start from (segment level), with the post hoc record-wise split on native signals. Vertical lines: range over the three seeds (no-overlap step and thinned-window control: one seed). Dotted line: chance for five balanced classes (published-cohort check: seven classes for Sharma-style, eight for Cheng et al.).}
\label{fig:waterfall}
\end{figure*}

With cohort, task and channels aligned and the record-wise split kept (step 2), all four methods still scored highly (Fig.~\ref{fig:waterfall}a). Mean segment-level balanced accuracy was 0.811 %
for Wadichar et al., 0.680 %
for the Sharma-style replication, 0.983 %
for Cheng et al.\ (window level) and 0.991 %
for Urtnasan et al. The lower Sharma-style value is not a drop at cohort alignment. At its published design, its epoch-level balanced accuracy was already 0.698%
, well below its accuracy. The method is trained without class balancing, so it favours the large classes, which accuracy rewards and balanced accuracy does not. Wadichar et al.\ balance the classes, and Cheng et al.\ and Urtnasan et al.\ are close to the ceiling, so for these three the two metrics barely differ. Removing Cheng et al.'s window overlap changed almost nothing (0.987%
; single seed).

Holding out whole patients removed most of this performance. Replacing the record-wise split by patient-wise cross-validation (step 3) lowered segment-level balanced accuracy in every replication and every seed. The means fell to 0.177%
, 0.315%
, 0.415 %
and 0.390%
, respectively. The mean drops were 0.634%
, 0.365%
, 0.568 %
and 0.601%
, and the smallest single-seed drop was 0.364%
. This is far above the margin set before the runs, so every replication lost performance once the test patients were unseen (largest Holm-corrected $p < 0.001$%
). For Urtnasan et al., this step compares a record-wise segment split, which deviates from the split the paper describes, with patient-wise cross-validation.

Wadichar et al.'s network fell below the chance level of one in five under patient-wise cross-validation. A post hoc, descriptive analysis of its run records traced this to the class balancing. Its code balances each class by keeping the first segments in subject order before any split. This retained segments of only 38 %
of the 94 %
patients. Under patient-wise folds, a class's share of the training segments then ranged from 7\,\% %
to 30\,\%%
, and per-class recall followed that share (Pearson $r$ = 0.38%
). This balancing therefore contributes to the below-chance result, and the Wadichar drop is the least clean of the four.

Harmonisation (step 4), still scored per segment, changed balanced accuracy little. It gave 0.175 %
for Wadichar et al.\ (patient-wise cross-validation 0.177%
), 0.324 %
for the Sharma-style replication (0.315%
), 0.371 %
for Cheng et al.\ (0.415%
) and 0.395 %
for Urtnasan et al.\ (0.390%
). The patient-level decision (step 5) then averaged the segment probabilities into one diagnosis per patient. The published methods reached patient-level balanced accuracies of 0.331 %
(Wadichar et al.), 0.332 %
(Sharma-style), 0.414 %
(Cheng et al.) and 0.447 %
(Urtnasan et al.). All were below the PL-LR benchmark of Section~\ref{sec:r-benchmark} (0.572%
); the comparison is descriptive, as signals, inputs and seeds differ.

Two leave-one-safeguard-out checks then identified which safeguard caused the loss. Both started from the harmonisation step, with every safeguard in place, and each undid exactly one of them. The record-wise-split check replaced the patient-wise folds by a single record-wise split of segments, so segments of every test patient were again present in training. The published-cohort check restored each paper's own cohort, task and channels, while keeping the patient-wise split and the harmonisation. If the loss comes from the split, the first check should recover the performance of cohort alignment and the second should not. Both checks were scored per segment, like the first four steps, and they located the loss in the split (Fig.~\ref{fig:waterfall}b; descriptive). Restoring only a record-wise split returned Cheng et al.\ (0.986%
) and Urtnasan et al.\ (0.991%
) to their aligned level. It recovered part of the gap for the Sharma-style replication (0.596%
) and Wadichar et al.\ (0.551%
). Keeping the published cohort and task with a patient-wise split gave values near or below the patient-wise step (0.190 %
to 0.342%
). Most of the loss therefore occurred at the split, not at cohort alignment, although the published-cohort check has more classes and so a lower chance level.

A post hoc check traced the incomplete recovery of the Sharma-style replication and Wadichar et al.\ to harmonisation. The record-wise-split check differs from cohort alignment in two ways: its inputs are harmonised, and its single split of segments replaces the published split. The post hoc check separated the two by applying the same record-wise split, with identical test segments, to the native signals of cohort alignment (Fig.~\ref{fig:waterfall}b). Both methods then returned to their cohort-alignment level: 0.672 %
for the Sharma-style replication (cohort alignment 0.680%
) and 0.823 %
for Wadichar et al.\ (cohort alignment 0.811%
). The change of split therefore explains little of the remaining gap, and harmonisation explains the rest. Under a record-wise split, part of these two methods' performance depended on recording properties that the harmonisation removes. Under patient-wise splits, the harmonisation changed little. Whether the same inflation occurs in this study's own models was tested next.

\begin{figure}[tb]
\centering
\includegraphics[width=\columnwidth]{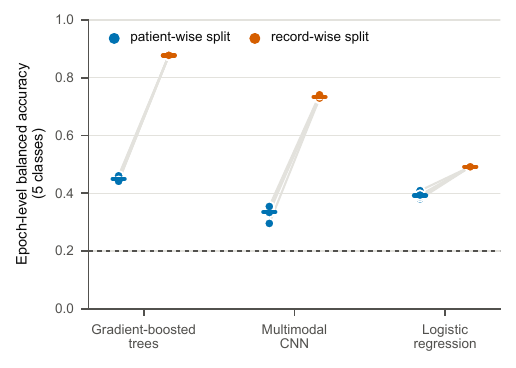}
\caption{Record-wise versus patient-wise splits of this study's models on identical inputs: 5-class epoch-level balanced accuracy (94 %
patients). Dots, seeds; lines join a seed's two results; bars, means; dashed line, chance for five balanced classes.}
\label{fig:inflation}
\end{figure}

\subsection{Record-wise inflation in this study's models}
\label{sec:r-inflation}

Record-wise splitting inflated this study's own models as well. Three of them were evaluated with both splits on identical inputs. Splitting epochs record-wise raised the 5-class epoch-level balanced accuracy in all three model pairs and every seed (Fig.~\ref{fig:inflation}). The gain was 0.427 %
for gradient-boosted trees (0.450 %
to 0.877%
), 0.398 %
for the multimodal CNN (0.335 %
to 0.733%
) and 0.100 %
for the epoch-level logistic regression (0.392 %
to 0.491%
); all paired $p < 0.001$%
. The two flexible models inflated far more than the linear model, consistent with their greater capacity to memorise recording-specific detail. This is why the benchmark of Section~\ref{sec:r-benchmark} was set under patient-wise evaluation only.

\begin{figure*}[t]
\centering
\includegraphics[width=\textwidth]{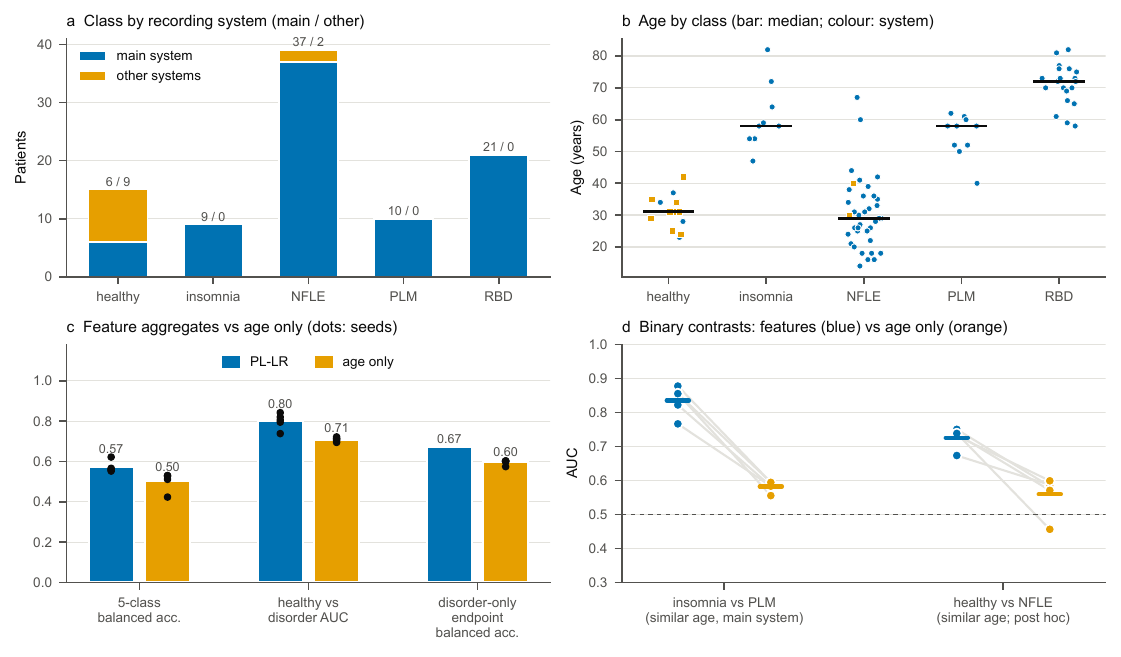}
\caption{Confounds at patient level (patient-wise 5-fold cross-validation). (a) Recording system by class. (b) Age by class; bar, median; colour, recording system. (c) Patient-level logistic regression (PL-LR) on the feature aggregates versus age only; dots, seeds (the PL-LR's disorder-only endpoint is a re-scoring of stored predictions, mean only). (d) Groups of similar age, features versus age only, per seed; insomnia versus PLM on the main system only (pre-specified), healthy versus NFLE on all systems (post hoc).}
\label{fig:confounds}
\end{figure*}

\subsection{Recording system, age and harmonisation}
\label{sec:r-confounds}

\begin{figure*}[t]
\centering
\includegraphics[width=\textwidth]{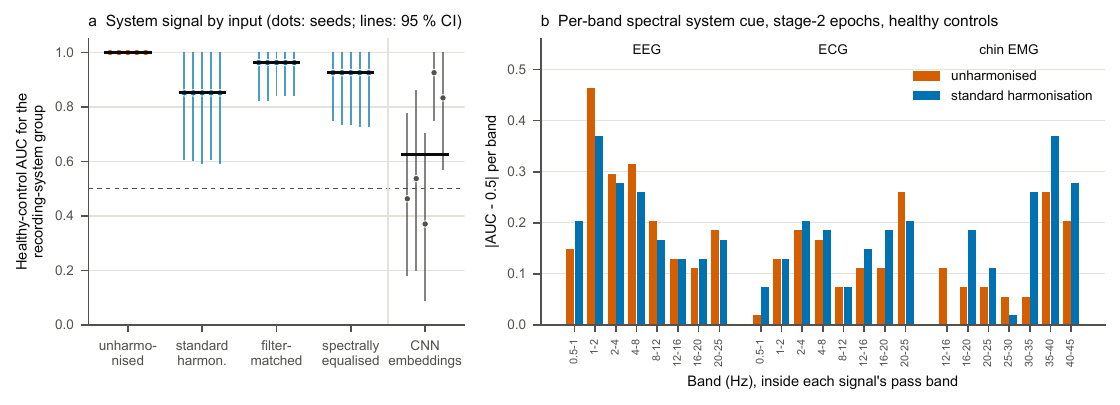}
\caption{Recording-system signal within the 15 %
healthy controls. (a) Healthy-control AUC for main versus other systems from unharmonised features, harmonised features (standard, filter-matched and spectrally equalised harmonisation) and the multimodal CNN's patient embeddings. Each dot is one seed, each vertical line its patient-level bootstrap confidence interval and each bar the mean over seeds. Feature-based AUCs are evaluated leave-one-out and so are identical across seeds; only their CIs vary. (b) Per-band system cue $|\mathrm{AUC}-0.5|$ of the stage-2 power spectrum before and after the standard harmonisation; one diagnostic run.}
\label{fig:system}
\end{figure*}

Recording system and class were strongly associated (Table~\ref{tab:cohort}, Fig.~\ref{fig:confounds}a). Of the 11 %
recordings from other systems, 9 %
were healthy controls and 2 %
were NFLE. Only 6 %
of the 15 %
healthy controls were recorded on the main system. Age also differed strongly between classes (Fig.~\ref{fig:confounds}b). The age ranges of healthy controls (23%
--42 %
years) and RBD patients (58%
--82 %
years) did not overlap. Insomnia and PLM had the same median age (58.0 %
years), and NFLE patients were of similar age to the healthy controls. Age was not an input of any model, but sleep signals change with age. Both recording system and age were therefore examined as possible explanations of the benchmark.

Harmonisation reduced the recording-system signal but did not remove it. Within the healthy controls, unharmonised features identified the recording-system group perfectly (AUC 1.000%
; Fig.~\ref{fig:system}a). The standard harmonisation reduced this to 0.852%
, with a lowest per-seed lower confidence bound of 0.589%
, so the signal remained detectable. Two stricter harmonisations were also tested. The filter-matched harmonisation added in software any hardware filter missing from a recording according to its header, and used steeper, narrower pass bands. The spectrally equalised harmonisation also divided each recording's spectrum by its own whole-night median spectrum. Both gave higher values than the standard harmonisation (AUC 0.963 %
and 0.926%
). Per frequency band, we measured the system cue in stage-2 epochs as $|\mathrm{AUC}-0.5|$. The largest within-band cue fell with harmonisation for EEG (0.46 %
to 0.37%
) and ECG (0.26 %
to 0.20%
), but rose for chin EMG (0.26 %
to 0.37%
; Fig.~\ref{fig:system}b). The CNN's patient embeddings separated the two systems within the healthy controls in 2 %
of 5 %
seeds (permutation test at the pre-specified significance level).

The system-aware views bound how much the healthy class rests on equipment. On main-system patients only, balanced accuracy was 0.530 %
for the PL-LR and 0.541 %
for the NSP; on the disorder-only endpoint it was 0.671 %
and 0.669%
. The lower 5-class values on main-system patients mean that part of the headline result rests on recognising the healthy controls of the other systems. Healthy recall was higher for other-system than for main-system healthy controls, by 0.344 %
for the PL-LR. This difference rests on only 6 %
and 9 %
healthy controls, however. Dropping features with a healthy-control system signal (the system-filtered candidate, C3) did not lower the system AUC of the outputs (difference 0.015%
, Holm $p = 1.000$%
).

Age, the second confound, came close to the benchmark on its own (Section~\ref{sec:r-benchmark}). Age alone reached a 5-class balanced accuracy of 0.504%
, a healthy-versus-disorder AUC of 0.709 %
and a disorder-only balanced accuracy of 0.596 %
(Table~\ref{tab:main}, Fig.~\ref{fig:confounds}c). Adding the feature aggregates to age raised the 5-class balanced accuracy by 0.067 %
(descriptive $p = 0.048$%
) and the disorder-only balanced accuracy by 0.068 %
($p = 0.062$%
). One pre-specified contrast was matched for both age and recording system: insomnia (9 %
patients) versus PLM (10 %
patients). Here the features reached AUC 0.836 %
against 0.582 %
for age (difference 0.253%
; Fig.~\ref{fig:confounds}d). This contrast shows signal beyond age, but it is small and descriptive. Healthy controls versus NFLE patients, of similar age but not on the same systems, gave 0.726 %
against 0.559 %
(post hoc). The benchmark is therefore above chance but close to what age alone achieves.

\begin{figure*}[t]
\centering
\includegraphics[width=\textwidth]{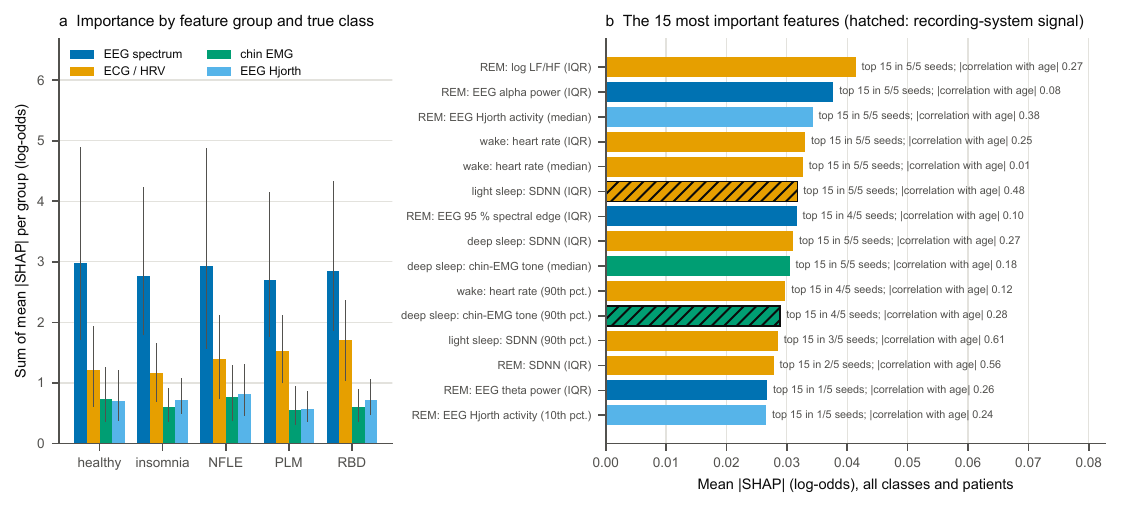}
\caption{Feature importance in the patient-level logistic regression (PL-LR): mean absolute SHAP value (exact linear SHAP, log-odds), each fold model explained on its own test patients. (a) Importance per feature group and true class; lines, range over seeds. (b) The fifteen most important features, coloured by group as in (a), with the number of seeds in which each was among the fifteen leading features and its absolute Spearman correlation with age; hatched, features that also separate the recording systems within the healthy controls (AUC above the cut-off of the system-filtered candidate C3, Section~\ref{sec:m-models}).}
\label{fig:shap}
\end{figure*}

\subsection{Feature attributions}
\label{sec:r-shap}

The SHAP explanations are compatible with the confound analysis. EEG spectral features carried the largest share of attribution (2.84 %
summed log-odds), followed by ECG features (1.40%
; Fig.~\ref{fig:shap}a). The EEG share reflects the size of that group, which has many more features than the ECG group. Individually, ECG and heart-rate-variability features were the most common among the fifteen leading features, which were mostly REM and wake summaries (Fig.~\ref{fig:shap}b). Of these fifteen, 2 %
carried a healthy-control recording-system signal above the cut-off of the system-filtered candidate. Their median absolute correlation with age was 0.26%
, reaching -0.61 %
for a light-sleep heart-rate-variability feature. Rankings were moderately stable between fold models (median Spearman correlation 0.54%
). Part of what the model uses may therefore reflect age or equipment rather than disorder-specific physiology.

\section{Discussion}
\label{sec:discussion}

Per patient, with every test patient unseen, the best result on this database was modest: above chance, but close to what age alone achieves. Little of the published performance survived patient-wise splits, and this study's own models inflated in the same way under record-wise splits. Recording system and age are entangled with diagnosis in this small single-centre cohort, and no external data were available. Even the benchmark is therefore an internal estimate that may still contain confounded signal.

Once patients were separated, model complexity did not improve performance detectably at this sample size. Handcrafted features helped the CNN, whereas the CNN alone did not match them. This suggests that with 94 %
patients the network learned less from the time--frequency images than the stage-wise feature aggregates provide. For signal-processing work on this database, the choice of evaluation therefore matters more than the choice of time--frequency representation.

The evaluation design largely explains why published results are so much higher. Record-wise splits inflate because the diagnosis is constant within a recording. Once segments of one night sit on both sides of a split, recognising the night is enough \cite{saeb2017,brookshire2024}. The three deep-learning methods dropped more than the feature-based Sharma-style replication. The leave-one-safeguard-out checks are consistent with this reading. Restoring only the record-wise split brought two methods back to their aligned level with every other safeguard in place, whereas restoring the published cohort and task did not. For the other two methods, the record-wise split restored their aligned level only on native signals (post hoc). Their record-wise advantage therefore depended partly on recording properties that the harmonisation removes. Recognition of recordings was not tested directly, however. The checks are descriptive, and part of the record-wise advantage may also come from neighbouring segments sharing transient physiology.

The same inflation occurred in this study's own models, so it is not specific to the published methods. On identical inputs, a record-wise split inflated even the linear model. It inflated the flexible models far more, as expected if capacity to memorise recording-specific detail drives the gain.

The confounds limit what any classifier can learn from this database. Harmonisation reduced the recording-system signal within the healthy controls but did not remove it, and in chin EMG the in-band cue even grew. Resampling, filtering and per-recording scaling cannot restore information that a lower sampling rate or a different hardware filter removed. Sampling rate and filters are both known to alter EEG and ECG measures \cite{widmann2015,ellis2015}; why the chin-EMG cue grew was not examined. Harmonisation was kept because no tested variant removed more. The disorder-only endpoint removes the recording-system confound but not age, and age alone came close to every endpoint. Age-related changes in sleep architecture, EEG power and heart-rate variability \cite{ohayon2004,carrier2001,umetani1998} overlap with the features the model relied on, as the SHAP correlations show. In pooled databases, too, disorders have been found more detectable in older than in younger subjects \cite{ali2026}. The insomnia-versus-PLM contrast, matched for age and recording system, shows signal beyond age, but it is small and descriptive.

\section{Limitations}
\label{sec:limitations}

The cohort is small: 94 %
patients from one centre, with one night each. The CIs are therefore wide, and the learning curve was still rising \cite{riley2020}. Recording system and age are entangled with diagnosis in this cohort, and no external data were available to separate them.

The replications depend on details we could not reproduce: the substituted filter bank, the narrower Cheng network, the inferred Urtnasan split and Wadichar et al.'s first-N class balancing. Where a re-implementation fell short of the published value, the original evaluation may have been leaky, or our re-implementation weaker. The measured drop compares the same model under two splits and does not depend on which. Statements about the published numbers do, in particular about Sharma et al.'s subject hold-out, which we did not reproduce. The step-wise attribution also assumes that step order does not matter, which the leave-one-safeguard-out checks test only in part.

Diagnoses and sleep stages were used as supplied, and sex was not analysed. Some harmonisation choices also remove physiological information. Clipping EEG at $\pm$50\,\textmu V flattens the largest slow waves of deep sleep, and the common 25-Hz upper band edge discards EEG activity above low beta. The 0.5--25\,Hz ECG band removes part of the QRS energy.

External validation on a larger multi-centre cohort is the most important next step, with age-matched healthy controls recorded on the patients' equipment \cite{steyerberg2016}. Published code would allow exact replications. Stating the split unit, the unit scored and the class set for every reported number would make CAP results comparable across studies.

\section{Conclusion}
\label{sec:conclusion}

In this study, sleep-disorder classification on the CAP Sleep Database was evaluated per patient, with every test patient unseen. A benchmark with one diagnosis per patient exceeded the empirical chance level but came close to an age-only model. Neither a multimodal CNN nor nested model selection improved on a logistic regression of stage-wise feature aggregates.

Four published methods were then re-evaluated under the same safeguards. Changing one safeguard at a time traced most of their near-perfect accuracy to the record-wise split. Their performance fell in every seed once whole patients were held out, and restoring only that split brought two of them back to their aligned level. This study's own models inflated in the same way when their epochs were split record-wise, the flexible models far more than the linear one.

Recording system and age are entangled with diagnosis in this database, and harmonisation reduced but did not remove the recording-system signal. The patient-wise benchmark and the safeguard protocol give future CAP classifiers a reference for comparison. Their value for diagnosis, however, can only be established on a larger multi-centre cohort with age-matched controls recorded on the patients' equipment.

\section*{CRediT authorship contribution statement}
\textbf{\mbox{Azwad Jahin Ifat:}} Conceptualization, Methodology, Software, Formal analysis, Visualization, Writing -- original draft. \textbf{\mbox{Md.\ Sajjad Hossain Mridul:}} Conceptualization, Methodology, Data curation, Investigation, Validation, Writing -- review \& editing. \textbf{\mbox{Mohammed Imamul Hassan Bhuiyan:}} Validation, Visualization, Supervision, Project administration, Writing -- review \& editing.

\section*{Data availability}
The CAP Sleep Database v1.0.0 is openly available from PhysioNet under the Open Data Commons Attribution License v1.0 \cite{capslpdb}.

\section*{Ethics statement}
This study used only the publicly available CAP Sleep Database, released through PhysioNet under an open licence \cite{capslpdb}. It is a secondary analysis: no new data were collected, the authors had no contact with the participants, and no further ethical approval was required.

\section*{Funding}
This research did not receive any specific grant from funding agencies in the public, commercial, or not-for-profit sectors.

\section*{Declaration of competing interest}
The authors declare that they have no known competing financial interests or personal relationships that could have appeared to influence the work reported in this paper.

\bibliographystyle{elsarticle-num}
\bibliography{references}

\end{document}